# Rethinking Digitalization and Climate: Don't Predict, Mitigate

Daria Gritsenko, Jon Aaen, and Bent Flyvbjerg

**Abstract**

Digitalization is a core component of the green transition. Today's focus is on quantifying and predicting the climate effects of digitalization through various life-cycle assessments and baseline scenario methodologies. Here we argue that this is a mistake. Most attempts at prediction are based on three implicit assumptions: (a) the digital carbon footprint can be quantified, (b) business-as-usual with episodic change leading to a new era of stability, and (c) investments in digitalization will be delivered within the cost, timeframe, and benefits described in their business cases. We problematize each assumption within the context of digitalization and argue that the digital carbon footprint is inherently unpredictable. We build on uncertainty literature to show that even if you cannot predict, you can still mitigate. On that basis, we propose to rethink practice on the digital carbon footprint from prediction to mitigation.

**1. Introduction**

Recently, debates on climate mitigation and adaptation have gained a new component: digital innovation. Digital technologies and infrastructures are expected to help in decarbonizing other industries and dematerializing consumption [1–4]. The European Union (EU) promotes the "Twin Transitions," proposing that green and digital goals complement each other well [5]. However, since the invention of modern computing and up till today, carbon emissions have risen continuously[1]. In other words, so far, the development and widespread diffusion of digitalization have not become a part of the solution to the climate crisis [7–9].

The growing interest in the links between digitalization and the climate crisis among practitioners and scholars has resulted in a vibrant discussion on how to calculate and predict the carbon footprint of digital technologies and initiatives most accurately. Scholars argue that accurate digital carbon footprint assessments are essential to avoid uncertainty and confusion, instigate action, and identify major reduction opportunities instead of minor reductions [10–12]. While this reasoning may seem compelling at first glance, we argue that this is a mistake. Most attempts at prediction are based on three implicit assumptions related to prediction: (a) digital carbon footprint can be quantified, (b) business-as-usual with episodic change leading to a new era of stability, and (c) investments in digitalization will be delivered within the cost, timeframe, and benefits described in their business cases. This paper shows that moving away from the attempts to make accurate predictions opens new opportunities to act climate-wisely regarding digitalization. We understand digital carbon footprint as the net sum of all GHG emitted in conjunction with deploying digital technology.

We start by reviewing the extant literature on the climate impact of digitalization, highlighting uncertainty as a key challenge in quantifying and predicting climate effects of digital technologies and initiatives. Building on decision-making under uncertainty frameworks [13,14], we propose an alternative perspective that focuses on mitigation rather than prediction of these effects. Concrete elements of our proposed solution are (1) transitioning into renewable energy as quickly and as effectively as possible (given that the primary negative climate impacts of digitalization are linked to increased (fuel-based) energy consumption); (2) organizing for continuous change with many interdependencies (given that digital technologies – and the whole digital ecosystem – continuously change in nonlinear ways), and (3) improving project delivery for digital-green initiatives (given that digital investments are performing significantly worse compared to other investment types in terms of cost overruns, delays, and benefit shortfalls). In doing so, this article explores how digitalization could be an effective part of the solution to the global climate crisis.

---

[1] Carbon emissions is not the only environmental effect of digitalisation. Other issues, including resource depletion, e-waste, and toxic substances, gained attention in the last decade due to rapid proliferation of digital devices and expanding scope of digitalization [6]. Yet, in this paper we limit the scope to considering carbon emissions only.

## 2. Understanding digital carbon footprint

The science of digitalization and climate contains a myriad of perspectives and approaches from various disciplines resulting in numerous, partially overlapping conceptualizations and inconsistent terminology. One dominant stream of literature distinguishes between *direct* and *indirect effects* [15–21]. Within this perspective, direct effects are always negative, referring to the environmental impacts of digital technologies related to material and energy consumption associated with the production, use, recycling, and disposal of digital devices and infrastructures. Indirect effects (sometimes termed 'induced effects') refer to a broad spectrum of both negative and positive climate effects following digitalization. Predictions of the climate effects of digitalization within this stream of literature can vary from including (a) direct effects only, or (b) include unwanted (increasing emissions) indirect effects, or (c) include all indirect effects on carbon emissions (both increasing and decreasing), or (d) 'net footprint', meaning include all GHG abatement minus (a)+(b) [22].

A second overarching stream of literature distinguishes between three levels of effects: *first-order*, *second-order*, and *other effects* (sometimes also called 'systemic', 'structural', or 'rebound'[2] effects) [24–29]. First order effects once again refer to the immediate effects of digitalization at the level of technology [30]. Two further levels seek to capture the difference between the second-order effects that are induced within the existing system (such as changes in production and consumption) and effects that potentially change the system itself. Positive second-order effects are expected to save energy and resources through dematerialisation and efficiency gains, while increase in labor and energy productivities are related to growth and may cause negative second-order effects [22]. Structural effects refer to the changes in economic structures, institutions, or culture, they mark behavioral shifts (positive, e.g., green consumerism, and negative, e.g., growth and re-materialization, [24], and they are commonly thought of as long-term, unintended, and far-reaching system-wide effects of introducing new digital technology [2]. E-commerce is a widely used example of how shopping from home reduces the need for shopping malls, changes logistics, and reduces travel (positive second-order effects). At the same time, it may lead to negative structural effects, such as increased consumption as the Internet is the world's largest advertising platform that simultaneously creates new desires (demand) and makes consumption easy (supply) [17].

---

[2] Some works refer to systemic level unintended effects as 'rebound effects', while others distinguish between different types of rebound effects, including direct rebound (or price effect) referring to rebound effects occurring on the same service, and indirect rebound referring to increase in other services due to improved efficiency (induction, income, and substitution effects)[23].

The frameworks discussed above form a basis to assess and quantify digital carbon footprint. Substantial efforts have been put into developing methodologies for evaluating the climate effects of digitalization [10,31,32]. Global and regional organizations such as the International Telecommunication Union (ITU), a United Nations specialized agency for ICT and ETSI (European Telecommunications Standards Institute) provide a number of standards and recommendations for assessing digital carbon footprint, including various life cycle assessments models of digital technologies, methods to assess the energy efficiency of digital infrastructures, and indicators (KPIs) to provide users of digital technologies with the tools to monitor their eco-efficiency and energy management. Bieser and Hilty[16] provided a review of other methodologies proposed by scholars, including Macro Economics [33], Scenario Analysis [34], System Dynamics [35], while newer approaches include Holistic Life Cycle Assessment (LCA) [36] and Higher-order effects of ICT in LCA[37].

Despite the differences, these methodologies face a number of similar challenges that have been aptly summarised by Bieser and Hilty [16]. The first challenge is baseline definition, which can either be fixed (at the moment when a digital technology is introduced or at the present date) or projection-based [38]. Carbon footprint estimations are highly sensitive to baseline definition, since predictions typically rely on business-as-usual scenarios as a reference point [39]. The second challenge is estimating the environmental impact. Typical approaches include top-down assessments (estimates of the overall energy consumption of the system divided by its components), bottom-up assessments (estimates energy use of each component through case-studies and combines these figures), and model-based assessments (model system components), and their estimates can differ by up to two orders of magnitude [19,40–42]. The final challenge is to predict the future, especially relating to the adoption of various systems [43,44] and future user behaviors [37,45], as well as interaction between the system components [46]. Which baseline is selected, how assessment is approached, and which assumptions about future use are taken has a significant impact on the results.

In sum, quantifying the climate effects of digitalization is not trivial. This is also reflected in how previous studies have yielded inconsistent results on the direct climate effects of digitalization [2,32], faced severe methodological challenges in attempting to calculate the indirect effects and systematically excluded structural effects from the calculations [47]. While conceptual frameworks are useful for understanding the complex link between digitalization and climate, we are skeptical about the fixation on trying to quantify and predict digital carbon footprint accurately. First, estimating the effects of digital technologies can be resource-intensive and require significant investments of time and money. Second, despite many attempts to calculate and predict the climate impact of digitalization, the inconsistent predictions provide little guidance to decision-makers as to what extent – and under which conditions – digitalization can alleviate the climate crisis [2]. Finally, if the information on the carbon footprint of a particular technology is inaccurate, efforts to reduce emissions from that technology may be misdirected or have unintended environmental consequences. Still, rather than factoring in uncertainty as an inherent trait of the complex relationship between digitalization and climate, most literature is devoted to improving the methods of calculations or expanding and refining available data. In what follows, we argue that the digital carbon footprint is inherently unpredictable. We draw on uncertainty literature to show that even if you cannot predict, you can still mitigate.

## 3. The climate impact of digitalization is not unknown – it is unknowable

The standard calculations of the climate effects of digital technologies are based on three implicit assumptions: (a) the digital carbon footprint can be quantified, (b) business-as-usual with episodic change leading to a new era of stability, and (c) investments in digitalization will be delivered within the cost, timeframe, and benefits described in their business cases. However, these assumptions are questionable, given uncertain inventory data and data gaps on digital technologies' energy consumption [48], the pervasive, dynamic, and open-ended nature of digital innovation [49], and the poor track record of digitalization projects performing significantly worse than expected with high risks of extreme budget overruns, delays, and benefit shortfalls [50]. Figure 1 presents a framework that summarizes how pervasive uncertainties result in an infinite variance of key variables, making it impossible to predict the climate effects of digitalization.

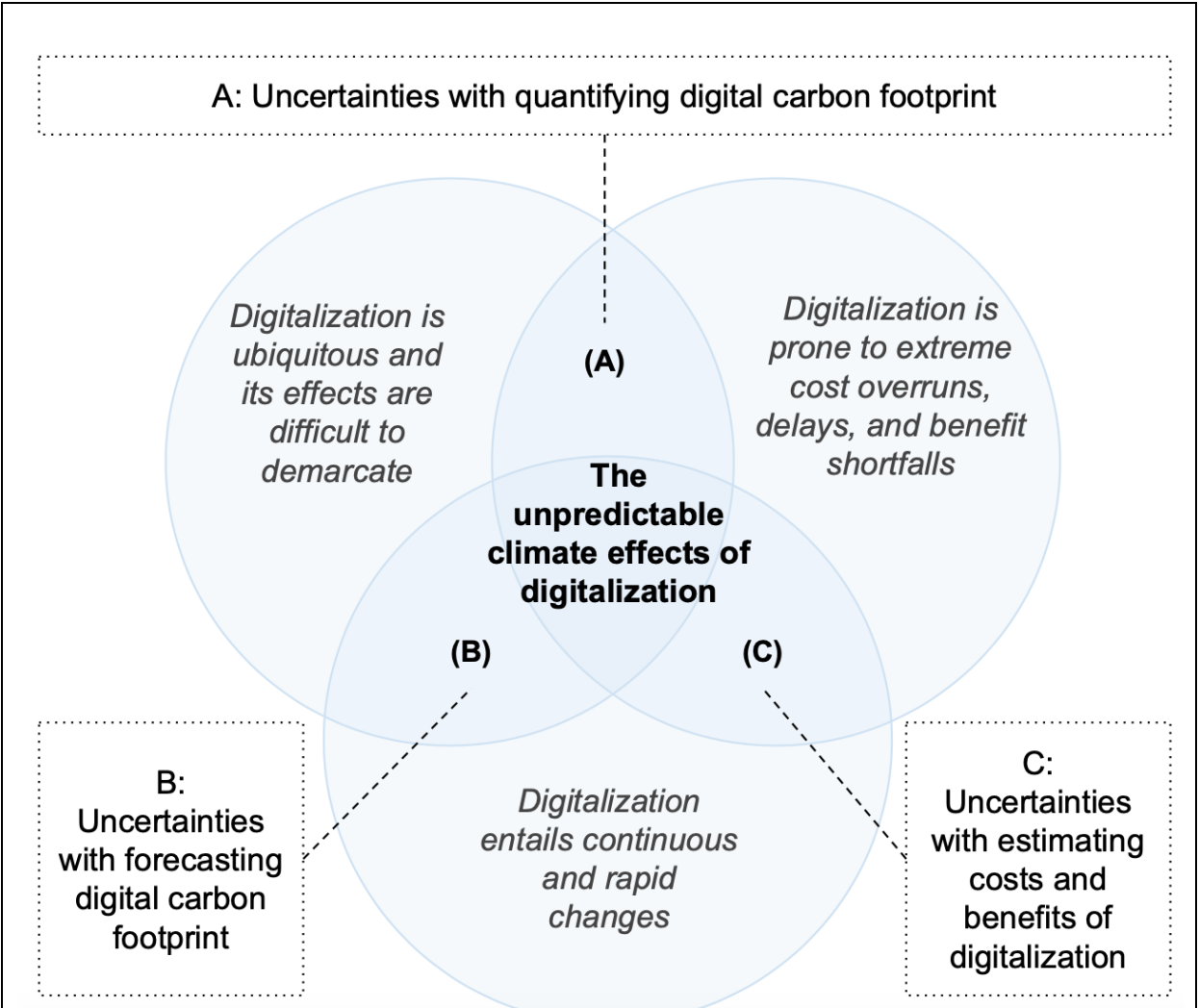

*Figure 1*. Uncertainties surrounding the predictability of digital carbon footprint. Source: Authors.

*3.1 Uncertainties with quantifying digital carbon footprint*

Even a reductionist approach to estimating the direct effects of digital technology is a cumbersome task that produces disputed and likely inaccurate information [2]. When the calculation aims to consider the entire lifecycle of digital technologies, a broad range of material processes should be included. Devices need to be manufactured, networks constructed, cables laid, data centers built and operated, etc., which makes it complex to demarcate the width and depth of climate impact. When supply chain pathways are chopped off, the climate impact becomes underestimated. Thus, estimates of digitalization carbon footprint are prone to truncation error, that is, partial exclusion of infrastructure by the traditional process of life-cycle assessment [2]. The complexity increases further if we consider the indirect and structural effects of digital technologies [51]. Digitalization brings about systemic changes with unforeseen consequences, such as increased consumption (goods and services) due to lower prices and efficiency gains (substitution effect) and changes in the overall framework in which agents make decisions (e.g., how information is produced, spread, and consumed) [52]. Studies that consider systemic changes of digitalization demonstrate that it is difficult – if not impossible - to quantify how digitalization impacts climate crises [30,47]. Even though the scope and extent of structural effects stemming from digitalization are impossible to quantify, we know that they constitute a significant problem [22,53].

### *3.2 Uncertainties with forecasting digital carbon footprint*

Quantified predictions rely on baseline scenarios with assumptions of business-as-usual, disrupted by infrequent episodic change that ends in a new era of stability [39,54,55]. However, such assumptions become particularly problematic in the context of digitalization since digital technologies – and their applications – continuously change in nonlinear ways, with new advancements being developed and implemented all the time. Research on digital innovation and transformation explains how the ability to continuously modify, combine, and scale up digital technologies creates generative and self-reinforcing mechanisms where technical malleability allows for the creation of new features, products, and services, attracting more users and partners that bring in more resources, expanding the reach of the system even further [56]. Thus, unlike traditional episodic change, where transformations are seen as distinct events with defined endpoints (e.g., Kurt Lewin's unfreeze-change-refreeze model, [57], digital innovations do not lead to stability; instead, they perpetuate continuous change [54]. This open-endedness makes digitalization a dynamic and rapidly changing phenomenon that is difficult to predict and keep track of, with even small changes having the potential to significantly impact various industries and aspects of daily life [49,58]. Data traffic is a case in point: growth in data demand is associated with increased energy demand, while at the same time since 2000 the electricity intensity of data transmission has decreased by half every two years [59]. Given the pace of technological innovation, many business-as-usual scenarios that seek to extend existing trends into the future seem obsolete and those rooted in current reality heavily on rely expert assessments which bring their own biases. Uncertainties related to users further complicate the projection-based baselines [37]. Social studies show that digitalization does not have a uniform, unidirectional effect on human behavior, but rather heterogeneous [60]. Though we expect that digitalization will impact human energy behavior, we do not know how stable these behaviors are, how many people will engage, and how they will further develop [52]. The ways how socio-technical systems develop are uncertain and that is why the total of all climate effects of digitalization includes too many parameters with uncertain values [53]. In other words, how digital carbon emissions will develop in the context of a rapidly advancing digital landscape is not only unknown – it is unknowable.

*3.3 Uncertainties with estimating costs and benefits of digitalization*

In addition to the complexities of estimating climate effects, digital investments are notorious for spiraling out of control, resulting in extreme budget overruns, delays, and benefit shortfalls. By analyzing a large sample of 5,392 digitalization projects, Flyvbjerg et al.[50] have empirically documented how digital investments are often managed so badly that performance measured the conventional way (cost, schedule, benefits) has infinite variance with an extremely fat tail[3], which means that when digitalization projects go wrong, they tend to go really wrong [50]. Almost one in five digitalization projects are delivered over budget by 50 percent or more – and the average overrun for those projects is a staggering 447 percent (Flyvbjerg & Gardner, 2023). Furthermore, fat tails make historical data a poor predictor of the future [14]. This means that we cannot apply traditional statistics based on variance and standard deviation as a basis for risk management of digitalization projects. Accordingly, the performance of digitalization projects is difficult to predict and involves a high risk of extreme cost overruns, delays, and benefit shortfalls. This adds an additional layer of uncertainty that needs to be considered when making decisions about digitalization and climate goals.

---

[3] The term "fat tails" refers to the behavior of the tails of a probability distribution. Representing the extreme values, a distribution's tail can be considered "fat" when the probabilities of extreme events are higher than what would be expected in a normal distribution. See Flyvbjerg et al. (2022) for the science of fat tails in IT projects.

Individually, each of the uncertainties described above is problematic for the predictability of digitalization and its climate impact. Taken together, they highlight that calculating the climate impact of digitalization is not merely a problem of methodology – it is a problem of uncertainty. Therefore, we need to move our perception of digital carbo footprint from the realm of 'unknown' (so we need to try harder to put a number on it) into the realm of 'unknowable' (too complex to model, historical data provide no useful guidance for future outcomes). By recognizing what is truly unknowable, we can avoid wasting time and resources trying to gather and estimate quantified information that is not available (or even falling into the false security of invented numbers) and instead focus on much-needed actions to mitigate rather than predict the climate impact of digitalization.

## 4. A way forward: Mitigating risks in the realm of the unknowable

If we cannot (reasonably) predict the carbon footprint of digitalization and its near-future developments, let alone accurately estimate the economic costs or potential benefits involved in specific digitalization projects, how should we make climate-wise decisions on digitalization? This question is critical to business managers and policymakers alike. Imagine, for instance, an organization that seeks to increase digital service delivery and requires additional infrastructural capacities, new IT products, etc., while remaining climate policy compliant. How should they plan and execute their digitalization initiatives in an environmentally responsible manner?

The good news is that you do not have to be able to predict something to be able to mitigate it. Recent literature on uncertainty – exemplified by Taleb et al. (2022), Sunstein (2021), and Kay and King (2020) – builds on the seminal work of Keynes (1921) and Knight (1921), and later Gumbel (1958) and Mandelbrot (1982). In essence, a distinction is made between 'risk', characterized by the predictability of a possibility of a future outcome, and 'uncertainty', characterized by both unknown future states and unknown consequences of actions. Decision-making under risk requires accurate data for calculating probabilities and optimizing the current course of action. Knightian or radical uncertainty requires decision-making beyond the numbers because reliable data are unavailable, so probabilities cannot be assigned to them. Instead, decision-making can be based on narratives (Kay & King, 2020), the maximin principle (Sunstein, 2021), the precautionary principle [14,61], or heuristics (Gigerenzer, 2022), to name a few prominent decision-making approaches 'beyond numbers'. While these approaches differ in some details, what they share at the base is adherence to guide decision-making under unquantifiable uncertainty. Below, we discuss how we can apply such approaches to instigate prudent decision-making about digitalization and the climate crisis in tandem.

**4.1 Renewable energy as the "tsunami wall" of digitalization's climate impact**

Knowing that something is "bad" without knowing "how bad" can support prudent decision-making as it constitutes practical knowledge [62]. Prudence means that we have a reasonable understanding of the uncertainties – even if we are uncertain about the magnitude, timing, or other specifics - and distinguish between identifying negative effects/consequences and assessing their severity. For instance, while tsunamis are unpredictable, their effects are known to be devastating. Tsunami walls and other countermeasures are used to mitigate the adverse effects of these natural catastrophes, independent of when they may happen and with which exact magnitude. The idea of limiting exposure is key to this type of precaution. The precautionary approach – dubbed in colloquial language 'better safe than sorry' – suggests that in situations with complex risks of severe and irreversible consequences, it is better to take preventive measures even if the effects and causal relationships are not fully established scientifically, rather than wait for proof of harm before acting (de Sadeleer, 2020).

Just as a tsunami wall is designed to protect coastal communities from the devastating effects of a tsunami, renewable energy can act as a protective wall against the adverse climate effects of digitalization. We already know that the negative climate impacts of digitalization are linked to increased (fuel-based) energy consumption due to the widespread use of devices, data centers, and other IT infrastructure [22]. Consequently, if digitalization were entirely based on renewables, the whole exercise of developing tools to quantify and estimate these links would become less relevant. Furthermore, if all digital devices and networks would run on renewable energy, it is a reasonable assumption that other sectors would also electrify and run on renewables. Thus, a widespread focus on renewable energy transition will also alleviate digitalization's indirect and rebound effects. In that case, the scope and scale of indirect effects would matter less.

Following this strategy suggests regulators to be stricter and impose requirements on data centers and transmission networks in terms of renewables (hence, 'green cloud' policies promoted by the EU aiming to make data centers carbon-neutral by 2030 are targeting the right spot, see EC, n.d.). For an organization that seeks to act climate-wisely in its digitalization efforts, this mitigation strategy translates into, e.g., installing onsite solar or wind power systems to generate electricity for their data centers and other digital infrastructure [63], choosing service providers that use renewable energy to power their servers [64], or using virtual machine migration techniques to intelligently transfer activities between placed geographically dispersed data centers to utilize renewable energy available elsewhere (Zhang et al., 2020).

**4.2 Organizing for a continuously changing digital landscape**

To account for the dynamic, interconnected, and open-ended nature of digitalization, digital technologies cannot be treated successfully as local, stand-alone solutions with a fixed end-point [54,65]. Instead, digital technologies should be understood as part of continuously evolving digital ecosystems unfolding in a complex interplay between multiple interconnected socio-technical components [56,58].

From the continuous change perspective, the climate effects of specific digital initiatives should be treated as part of an ongoing process, requiring continuous considerations and adjustments to align with the ongoing, evolving and cumulative change induced by digitalization [54]. Accordingly, for organizations to effectively manage their digital carbon footprint, they must move away from traditional change management perceptions of episodic change that ends in a new era of stability in favor of continuous change perspective [55]. The shift from episodic to continuous change management compels organizations and policy-makers to adopt adaptive and responsive strategies when considering their digitalization efforts and climate effects in tandem.

Continuous change perspective also draws attention to the interconnectedness and interdependence of digital technologies. On the one hand, change in one component is likely to generate change in other parts of the digital ecosystem [56], on the other hand, a single change agent has a limited power to dictate the processes and outcomes of digital innovation [66–68]. As a result, managing the digital carbon footprint requires organizations to consider not only the effects of their digitalization initiative on a local micro-level but also the macro-level impact on the digital ecosystem. This requires new approaches to organize innovation efforts and engage with actors beyond organizational boundaries, involving coordination with a diverse set of actors and external partners [49,68].

Organizations should, therefore, consider both micro-level and macro-level effects by zooming in to assess the specific climate impacts of individual digital initiatives and simultaneously zooming out to understand the broader context of how these individual components contribute to the overall climate effects within the digital ecosystem.

### 4.3 Cutting the tail of extreme cost overruns in digitalization projects

Whereas renewable energy projects (solar, wind, water) generally perform well when it comes to being delivered on time, within budget, and providing the expected benefits – digitalization projects are infamous for the opposite (Flyvbjerg & Gardner, 2023). Thus, beyond transitioning to renewable energy sources, how to deliver digitalization effectively and ethically will be the difficult remaining question to address to fully utilize digital technologies as a core part of the solution to the climate crisis.

Poor project performance in digitalization can hamper organizations' ability to realize climate goals in several ways. Firstly, poor project performance can have an obvious negative impact on an organization's ability to achieve a positive return on investment (ROI) and may even lead organizations into bankruptcy if large digitalization projects spin out of control[50]. Secondly, when digitalization projects run over budget or schedule, they can divert resources away from other initiatives or result in the postponement or cancellation of projects, as showcased in the infamous TAURUS project that wasted 11 years of development, exceeding the initial budget with 13.200%, and still failed to deliver a paperless digital solution for the London Stock Exchange [69]. Thirdly, when digitalization projects experience delays or benefits shortfalls, it can result in missed opportunities to implement green technologies or initiatives. For example, if a project to implement energy-efficient hardware and software is delayed, it can result in the continued use of older, less efficient technology and processes[70]. Moreover, previous failures, project delays, or benefits shortfalls can also lead to reduced stakeholder support and decreased motivation to engage in future projects[71].

Accordingly, for organizations to act climate-wisely regarding their digitalization efforts, they must mitigate the adverse effects of digitalization projects spinning out of control. When faced with extreme fat-tailed risks, decision-makers should steer away from traditional risk management approaches (based on assumptions of normal distribution) as they give a false sense of security by systematically underestimating the potential hazards involved in digitalization projects[50]. Instead, decision-makers should focus on either avoiding high-risk projects entirely (i.e., more careful selection of which digitalization projects to engage in (see [72] or "cutting the tail" in order to reduce the likelihood or severity of digitalization projects spinning out of control. As suggested by Flyvbjerg et al[50], cutting the tail in digitalization means that managers must pay particular attention to how digitalization projects often consist of interconnected components (software, sensors, and communication devices), where a problem in a single component can lead to chain reactions affecting other connected components. Therefore, organizations need to identify the critical components, allocate additional resources to manage them proactively, and – if possible – reduce the number of interdependencies in project delivery.

## 5. Concluding remarks

The goal of this paper was to give an overview of the extant literature on digitalization and the climate crisis followed by a suggested reset – new ways of thinking about digitalization and climate in tandem. We have shown that current literature is focused on identifying and quantifying the links between digitalization and climate crisis by attempting to quantify and predict the climate impact of digitalization. While we appreciate the extensive work to conceptualize various types of climate effects, we show that these assessments, within the context of digitalization, are unavoidably linked to uncertainty. We argue that to support strategic decision-making for policy-makers and business managers who are willing to make climate-wise decisions on digitalization, this uncertainty needs to be taken seriously.

**Data availability statement:** We do not analyse or generate any datasets.